\documentstyle[aps,prbbib,twocolumn,epsf]{revtex}

\begin{document}
\draft
\title{Parametric quantum spin pump}

\author{Wei Zheng$^1$, Junling Wu$^2$, Baigeng Wang$^{2,3}$, Jian 
Wang$^{2,1,a}$, Qingfeng Sun$^4$, and Hong Guo$^4$}
\address{1. Institute of Solid State Physics, Chinese Academy of
Sciences, Hefei, Anhui, P. R. China\\
2. Department of Physics, The University of Hong Kong, 
Pokfulam Road, Hong Kong, China\\
3. National Laboratory of Solid State Microstructures and 
Department of Physics, Nanjing University, Nanjing, P.R. China\\
4. Department of Physics, McGill University, Montreal, PQ, H3A, 2T8,
Canada\\
}
\maketitle

\begin{abstract}

We investigate a non-adiabatic parametric quantum pump consists of a 
nonmagnetic scattering region connected by two ferromagnetic leads. 
The presence of ferromagnetic leads allows electrons with different spins
to experience different potential landscape. Using this effect we propose 
a quantum spin pump that drives spin-up electrons to flow in one direction 
and spin-down electrons to flow in opposite direction. As a result, the 
spin pump can deliver a spin current with vanishing charge current.

\end{abstract}



A parametric quantum pump generates a DC electric current by a cyclic
variation of system parameters while keeping the leads at a constant 
chemical potential\cite{switkes}. Considerable effort has been devoted 
to understand the physics of parametric 
pumping\cite{brouwer,zhou,wei1,shutenko,avron,levinson,buttiker}.
It was found that the pumped current is rather sensitive to various 
parameters of the system such as potential of the pump, frequency of the
driving force, and Fermi energy of the leads. As parameters vary, the
pumped current can even change directions: this has been predicted 
for charge pumping in resonant tunneling diodes\cite{wei1,buttiker1}, 
nanotube quantum pumps\cite{wei2}, finite frequency pumping 
process\cite{wbg1}, and quantization of the pumped charge\cite{aharony}.  
To fully exploit this behavior, in this paper we investigate parametric 
pumping in the presence of ferromagnetic leads. When ferromagnetic
leads are present, electrons with different spins experience different 
potentials. Under certain conditions the pumped current for electron with 
different spin may flow in opposite directions. As a result, the total 
pumped charge current can be zero while a pure spin current is delivered.
Such a quantum spin pump may have potential applications in the 
fascinating field of spintronics\cite{all}.

Recently, several different spin pumps have been proposed. In the uni-pole 
spin battery studied in Refs.\onlinecite{brataas,wbg3}, a spin current is 
generated by either a rotating magnetic moment or rotating external magnetic 
field. In these uni-pole pumps, spin current is not conserved due to spin
flips\cite{brataas,wbg3}. In contrast, the spin pump considered in this
work satisfies spin current conservation when magnetizations of the
two leads are parallel or antiparallel. In another direction, an adiabatic 
quantum pump was proposed\cite{chamon} such that a spin polarized current is
generated in a chaotic quantum dot in the presence of an in-plane magnetic 
field.  In this work we go beyond the adiabatic regime and examine the
frequency dependence of the pumped spin current. Our results show that a
pure spin current can be achieved by varying the pumping frequency.

The system we investigate is a spin-valve which consists of a non-magnetic 
scattering region connected by two ferromagnetic electrodes to the reservoir. 
The magnetic moment ${\bf M}$ of the left electrode is pointing to the 
$z$-direction, the electric current is flowing in the $y$-direction, while 
the moment of the right electrode is at an angle $\theta$ to the $z$-axis 
in the $x-z$ plane. For simplicity of discussion, we assume that the value of 
molecular field $M$ is the same for the two electrodes, thus a standard
spin-valve effect is obtained\cite{slon} by varying the angle $\theta$. 
Essentially, $M$ mimics the difference of density of states (DOS) between 
spin-up and down electrons\cite{slon} in the electrodes. The pumped current 
for this system can be calculated at finite frequency and up to the second 
order in pumping amplitude using the perturbation 
theory based on nonequilibrium Green's functions\cite{wbg1}. 
The pumped particle current due to spin component $\beta$ through lead
$\alpha$ is found to be ($\hbar=1$)\cite{wu}
\begin{eqnarray}
&&J_{\alpha \beta}=i \int \frac{dE}{2\pi} {\rm Tr} 
\left[{\bf \Gamma}_\alpha {\bf G}^r 
[(f_{-}-f) V ({\bf G}^r(E_{-})
-{\bf G}^a(E_{-})) V^* \right. \nonumber \\
&&\left.+ V^* (f_+-f) ({\bf G}^r(E_+)- 
{\bf G}^a(E_+)) V ] {\bf G}^a \right]_{\beta \beta} 
\label{eq8}
\end{eqnarray}
where $\beta = \uparrow,\downarrow$, $V$ is the pumping potential 
profile\cite{wu}, $E_{\pm}=E \pm \omega$, and $G^r$ is a $2 \times 2$ 
matrix representing the equilibrium retarded Green's function
\begin{equation}
{\bf G}^r(E)=\frac{1}{E-H_0-{\bf \Sigma }^{r}}
\end{equation}
with $H_0$ the Hamiltonian in the absence of pumping potential. 
Here ${\bf \Sigma}^{r}\equiv \sum_{\alpha} {\bf \Sigma}_{\alpha }^{r}$ 
is the self energy and ${\bf \Gamma}_{\alpha}=
-2Im[{\bf \Sigma}_{\alpha}^{r}]$ is the linewidth function. 
The self-energies are given\cite{wbg2}

\begin{equation}
{\bf \Sigma}^r_\alpha(E) = \hat{R}_\alpha \left( 
\begin{array}{cc}
\Sigma^r_{\alpha\uparrow} & 0 \\ 
0 & \Sigma^r_{\alpha\downarrow}
\end{array}
\right) \hat{R}^\dagger_\alpha
\label{self}
\end{equation}
with the rotational matrix $\hat{R}_\alpha$ for electrode $\alpha$ 
defined as 
\begin{equation}
\hat{R} = \left( 
\begin{array}{cc}
\cos\theta_\alpha/2 ~~ & \sin\theta_\alpha/2 \\ 
-\sin\theta_\alpha/2 ~~ & \cos\theta_\alpha/2
\end{array}
\right)\ \ .
\end{equation}
Here angle $\theta_{\alpha}$ is defined as $\theta_L=0$ and 
$\theta_R=\theta$ and $\Sigma^r_{\alpha \sigma}$ is the usual
self energy\cite{jauho}. In this paper, we will study two special cases:
$\theta=0$ or $\pi$ for spin current. From Eq.(\ref{eq8}), we observe
that up to the second order in pumping amplitude, the particle can
absorb or emit a photon during the pumping process. The contribution due
to these two photon assisted processes have different sign and tend to
cancel each other. As the result of this competition, the pumped 
particle current can reverse its direction upon varying system parameters. 

The pumped charge current $I_\alpha$ is given by
\begin{equation}
I_{\alpha}=q ~ (J_{\alpha \uparrow}+ J_{\alpha \downarrow})
\label{ele}
\end{equation}
and the pumped spin current $I_{s\alpha}$ is (we have set $\hbar=1$)
\begin{equation}
I_{s\alpha}=(J_{\alpha \uparrow}- J_{\alpha \downarrow})/2\ .
\label{spin}
\end{equation}
Now we examine the conservation law for pumped current. When
$\theta=0,\pi$, the $2 \times 2$ Green's function ${\bf G}^r$ and self
energy are diagonal. As a result, we have 
\begin{eqnarray}
&&\sum_\alpha J_{\alpha \beta}=i \int \frac{dE}{2\pi} {\rm Tr} 
\left[{\bf G}^a {\bf \Gamma} {\bf G}^r \right. \nonumber \\
&& \times \left. (f_{-}-f) V ({\bf G}^r(E_{-})
-{\bf G}^a(E_{-})) V^* \right. \nonumber \\
&&\left.+ {\bf G}^a {\bf \Gamma} {\bf G}^r 
(f_+-f) V^* ({\bf G}^r(E_+)- 
{\bf G}^a(E_+)) V  \right]_{\beta \beta} 
\label{eq9}
\end{eqnarray}
where we have moved ${\bf G}^a$ to the beginning of the trace. This can
be done only if the Green's function and self energy are diagonal. Using
the fact that ${\bf G}^r-{\bf G}^a = -i {\bf G}^a {\bf \Gamma} {\bf
G}^r$ and changing variable from $E_+$ to $E$ in the second term of
Eq.(\ref{eq9}), it is straightforward to show the $\sum_\alpha J_{\alpha
\beta}=0$. This means that both the pumped electric current and pumped
spin current are conserved in this device.

Now we use Eqs.(\ref{eq8}), (\ref{ele}), and (\ref{spin}) to calculate 
pumped electric current and spin current.
The system we studied is a symmetric double $\delta$ barrier structure 
modeled by potential $U(x)=V_0 \delta (x+a)+V_0 \delta (x-a)$.
For this system the Green's function $G(x,x')$ 
can be calculated exactly\cite{yip}. 
The pumping potential is chosen to be sinusoidal $V(x,t)=V_p [\cos(\omega t) 
\delta(x-x_1)+ \cos(\omega t+\phi) \delta(x-x_2)]$. 
We will calculate the pumped electric and spin current from 
the left lead at zero temperature and set $V_0=69.7$. We set
$x_1=-a$ and $x_2=a/10$. A gate voltage $v_g$ is applied in the double 
barrier structure to control the resonant electron level. We assume that
the Fermi level of the leads is in line with the resonant level at
$v_g=0$. Finally the unit is set by $\hbar=2m=q=2a=1$.  For the 
system of Fe/Ge/Fe with $a=1000\AA$, the energy unit is $E=0.046meV$
which corresponds to $\omega=11.0$ GHz. The unit for pumped current is
$2\times 10^{-7}$A. The unit for spin current is $2 meV$. 

In Fig.1, we show the pumped electric current (solid line) and spin 
current (dotted line) as a function pumping frequency when $\theta=0$
for fixed Fermi energy and magnetization $M$. As the frequency increases, 
the pumped electric current increases and reaches a maximum 
value at $\omega=0.0077$. As the frequency increases further, the pumped
electric current decrease and becomes negative at large frequency. 
In the presence of magnetic electrodes, the pumped electric current is spin 
polarized with non-zero spin current. The behavior of the pumped spin
current is similar to that of the pumped electric current. It is 
positive at small pumping frequency and becomes negative at
larger frequencies. At $\omega=0.0168$ (thin vertical line in Fig.1), 
the pumped electric current is zero and a pure spin current
is achieved. The physics behind this is the following.
During parametric pumping, the system pumps out spin-up and spin-down 
electrons. For electron with a given Fermi energy, the potential of the 
ferromagnetic lead for spin-up electron is $-M$ while for spin-down electron 
it is $M$. As a result, the pumped electric current for different spins can 
have different sign, {\it i.e.} spin-up electron pumps out from the right 
lead to the left whereas spin-down electron flows to the right lead during
the pumping. This way a spin current is generated. At certain value of 
the frequency, a complete cancellation of electric current occurs and
a pure spin current is delivered.

In Fig.2 we plot the pumped electric current and spin current versus
pumping frequency for $\theta=\pi$. We observe that the pumped electric
current for $\theta=0$ has the same order of magnitude as that at 
$\theta=\pi$. The pumped spin current, however, is quite different from 
that at $\theta=0$. It displays a linear dependency on the pumping frequency. 
The frequency at which a pure spin current occurs is also different from the 
case when $\theta=0$. Comparing with the case of $\theta=0$, the spin 
current reverses its direction when $\theta=\pi$. 
Our study shows that the pure spin current is
a generic property and can occur at a wide range of parameters. In
Fig.1, we see that at $M=2$ and $\omega=0.0168$ the system pumps out
pure spin current. As we vary $M$, the frequency $\omega$ at which the
pure spin current occurs also changes. The trajectory or the 
"phase diagram" is depicted in Fig.3 where pumping frequency versus 
magnetization of the leads for $\theta=0$ is shown (solid line) along
with the magnitude of pumped spin current. 

In summary, we have proposed a non-adiabatic quantum spin pump which can 
generate a pure spin current with zero electric current during the parametric 
pumping. The device consists of a non-magnetic system connected by two 
ferromagnetic leads. Since electrons with different spin experience
different potential landscape, the corresponding pumped current 
can be quite different. At certain condition, the pumped electric
current can be zero while pure spin current is produced. 
Our numerical results show that this can be easily achieved by varying 
system parameters such as the pumping frequency.

\section*{Acknowledgments}
We gratefully acknowledge support by a RGC grant from the SAR Government of 
Hong Kong under grant number HKU 7113/02P and from NSERC of Canada and FCAR 
of Quebec (H.G).

\bigskip
\bigskip
\bigskip
\noindent{$^{a)}$ Electronic mail: jianwang@hkusub.hku.hk}

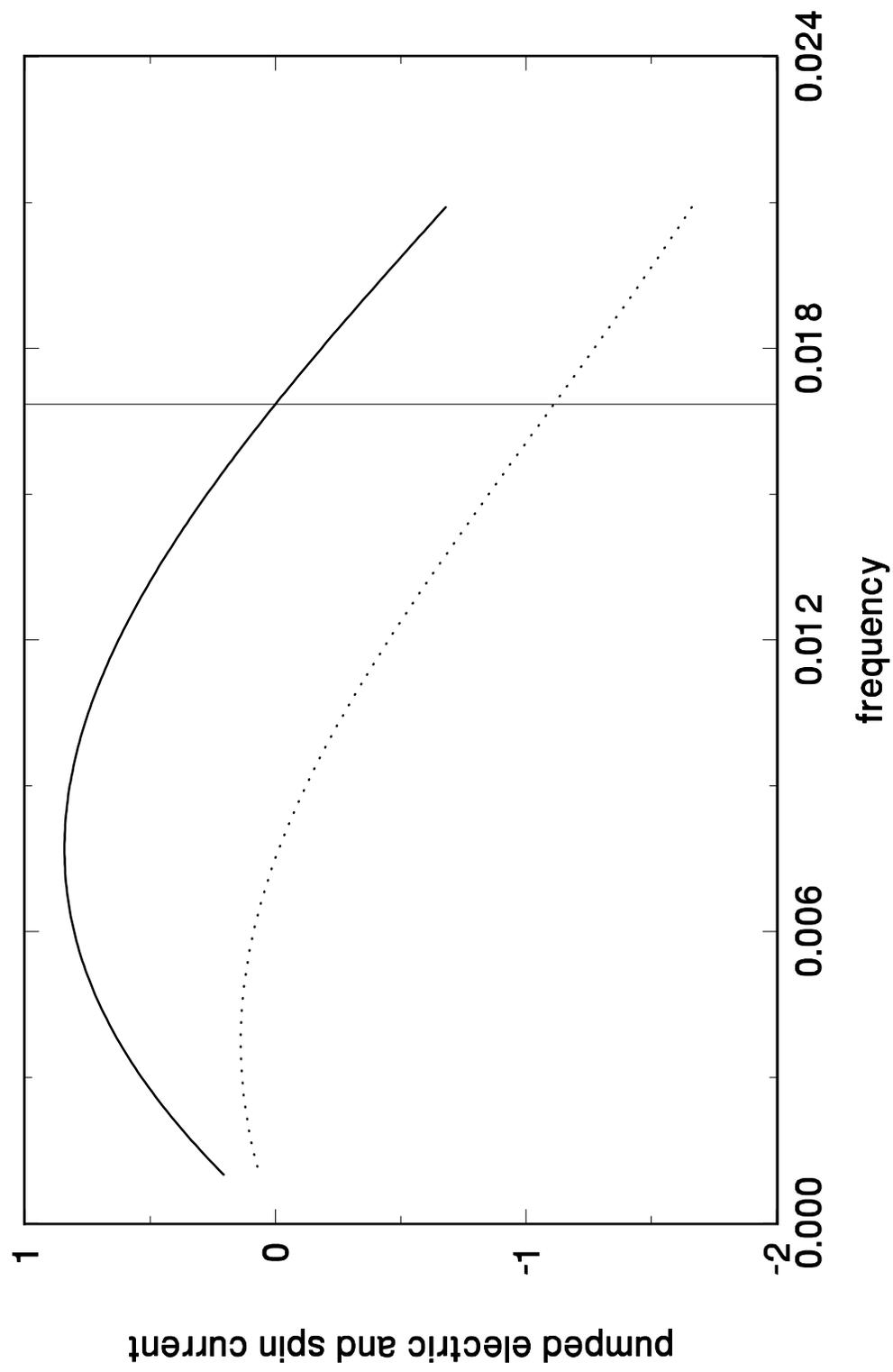
\begin{figure}
\caption{
The pumped electric current (solid line) and pumped spin current (dotted
line) as a function of pumping frequency at $\theta=0$. The pumped
current for spin-up and down electrons (dot-dashed line and long-dashed
line) are also plotted. Here $M=2$ and $E_F=3.4$. The vertical thin line
corresponds to the frequency at which the pumped electric current is
zero.
}
\end{figure}

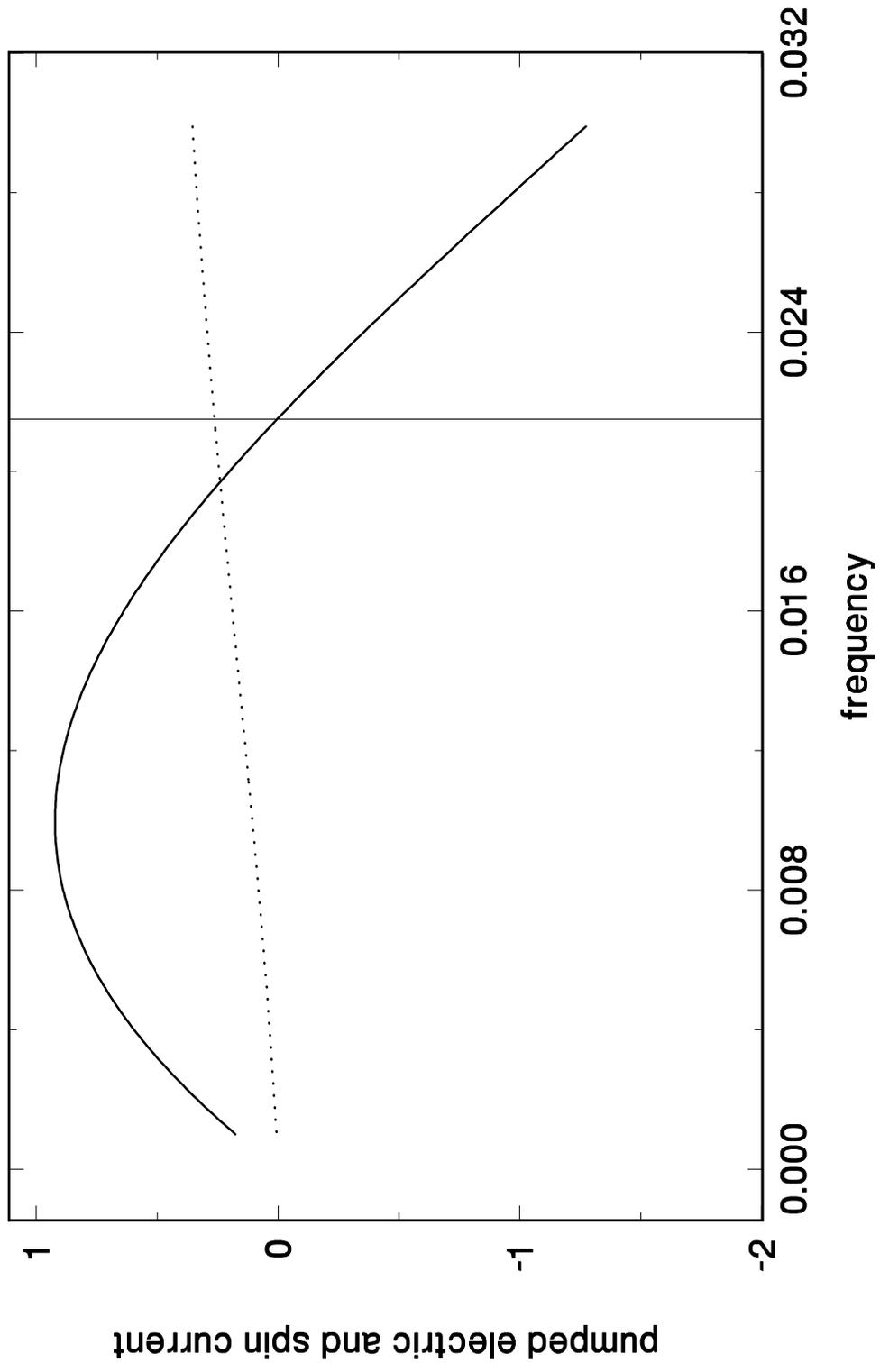
\begin{figure}
\caption{
The pumped electric current (solid line) and pumped spin current (dotted
line) as a function of pumping frequency at $\theta=\pi$. Other
parameters are the same as in Fig.1.
}
\end{figure}

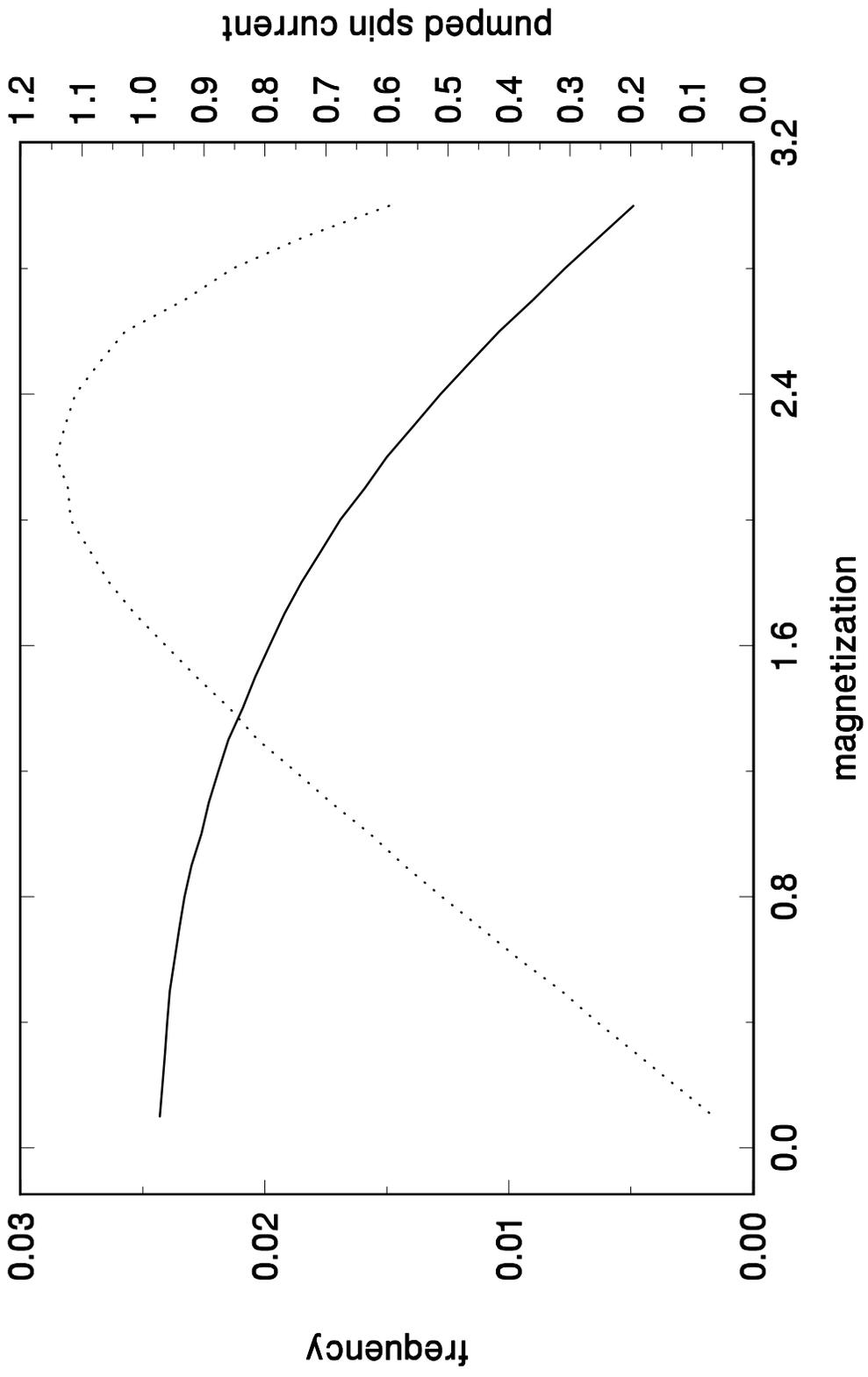
\begin{figure}
\caption{
The pumping frequency (solid line) and pumped spin current (dotted line)
versus magnetization of the ferromagnetic lead at which the pumped
electric current is zero. 
}
\end{figure}


\begin{thebibliography}{00}
\bibitem{foot2}
For the GaAs system with $a=1000A$, the energy uint is $E=56 \mu eV$.
\end{thebibliography}
\end{document}